\documentclass{aa}              
\usepackage{graphics}

\newcommand\kms{$\mbox{km\,s}^{-1}$}
\newcommand\mpc{$h^{-1}\mbox{Mpc}$}
\newcommand\der{\mbox{d}}
\newcommand\di{\delta_i}
\newcommand\loc{{\mathrm l}}
\newcommand\vmean{\overline{v}}
\newcommand\tableheadone{
\begin{flushleft}
\begin{tabular}{lr@{$^{\mathrm h}$}@{\extracolsep{0pt}}l@{$^{\mathrm m}$}l@{,}r@{$\degr$}l@{$\arcmin$}l@{$\arcsec$\hspace{12pt}}@{\extracolsep{-2pt}}rrcr}
\hline
\noalign{\smallskip}
\multicolumn{1}{c}{Cluster}
   &\multicolumn{6}{c}{Barycenter coords.}
   &\multicolumn{1}{c}{$e$} 
   &\multicolumn{1}{c}{$\theta$}
   &\multicolumn{1}{c}{$r_{\mathrm{eq}}$}
   &\multicolumn{1}{c}{$N_{\mathrm c}$}\\
&\multicolumn{6}{c}{RA (B1950) Dec} 
   & &\multicolumn{1}{c}{(rad)} &\multicolumn{1}{c}{(\mpc)}&\\
\multicolumn{1}{c}{(1)}
   &\multicolumn{6}{c}{(2)}
   &\multicolumn{1}{c}{(3)}
   &\multicolumn{1}{c}{(4)} &\multicolumn{1}{c}{(5)} 
   &\multicolumn{1}{c}{(6)}\\
\noalign{\smallskip}
\hline
\noalign{\smallskip}
}
\newcommand\tableheadtwo{
\begin{flushleft}
\begin{tabular}{lrccr}
\hline
\noalign{\smallskip}
\multicolumn{1}{c}{Cluster}
   &\multicolumn{1}{c}{$N$}  
   & \multicolumn{1}{c}{$p(S)$}
   & \multicolumn{1}{c}{$p(K)$}
   & \multicolumn{1}{c}{$p(\Delta)$}\\
\multicolumn{1}{c}{(1)} &\multicolumn{1}{c}{(2)} &\multicolumn{1}{c}{(3)} 
	&\multicolumn{1}{c}{(4)} &\multicolumn{1}{c}{(5)}\\
\noalign{\smallskip}
\hline
\noalign{\smallskip}
}
\newcommand{\aj}{AJ}         
\newcommand{\aaa}{A\&A}      
\newcommand{\aas}{A\&AS}     
\newcommand{\apj}{ApJ}       
\newcommand{\apjs}{ApJS}     
\newcommand{\mnras}{MNRAS}   
\newcommand{\pasp}{PASP}     

\begin{document}
\thesaurus{03(03.13.02; 11.03.1; 12.03.3)}
\title{Substructure in the ENACS clusters} 

\author{Jos\'e M.\,Solanes\inst{1} \and Eduard Salvador-Sol\'e\inst{2} \and
 Guillermo Gonz\'alez-Casado\inst{3}} 

\institute{Departament d'Enginyeria Inform\`atica, Universitat Rovira i
 Virgili. Carretera de Salou, s/n; E--43006~Ta\-rra\-go\-na, Spain.\and
 Departament d'Astronomia i Meteorologia, Universitat de
 Bar\-ce\-lo\-na. Av.~Diagonal 647; E--08028~Barcelona, Spain.\and
 Departament de Matem\`atica Aplicada II, Universitat Polit\`ecnica de
 Catalunya. Pau Gargallo 5, E--08028~Bar\-ce\-lo\-na, Spain.\\E-mail:
 solanes@etse.urv.es, eduard@faess0.am.ub.es, guille@coma.upc.es}

\offprints{J.M.\,Solanes}

\date{Received $\ldots$ / Accepted $\ldots$ }


\maketitle


\begin{abstract}
Subclustering is investigated in a set of 67 rich cluster galaxy
samples extracted from the ESO Nearby Abell Cluster Survey (ENACS)
catalog. We apply four well-known statistical techniques to evaluate
the frequency with which substructure occurs. These diagnostics are
sensitive to different aspects of the spatial and velocity distribution
of galaxies and explore different scales, thus providing complementary
tests of subclustering. The skewness and kurtosis of the global radial
velocity distributions, useful for judging the normality, and the
powerful $\Delta$ test of Dressler \& Shectman, which measures local
deviations from the global kinematics, show that the ENACS clusters
exhibit a degree of clumpiness in reasonable agreement with that found
in other less homogeneous and smaller cluster datasets. On the other
hand, the average two-point correlation function of the projected
galaxy distributions reveals that only $\sim 10\%$ of the systems
investigated show evidence for substructure at scale lengths smaller
than 0.2\,\mpc. This is much less than in earlier studies based on the
Dressler \& Shectman's cluster sample. We find indications of a
possible systematic deficiency of galaxies at small intergalactic
separations in the ENACS clusters.
\end{abstract}


\section{Introduction}
In the last two decades considerable attention has been focused on the
study of substructure within rich clusters of galaxies. The importance
of subclustering lies in the information it conveys on the properties
and dynamics of these systems, which has chief implications for
theories of structure formation. A number of authors have developed and
applied a variety of methods to evaluate the clumpiness of galaxy
clusters both in the optical and X-ray domains (e.g. Geller \& Beers
1982\cite{GB82}; Fitchett \& Webster 1987\cite{FW87}; West et
al. 1988\cite{We88}; Dressler \& Shectman 1988a\cite{DS88a}, hereafter
DS88; West \& Bothun 1990\cite{WB90}; Rhee et al. 1991\cite{Rh91};
Jones \& Forman 1992\cite{JF92}; Mohr et al. 1993\cite{Mo93};
Salvador-Sol\'e et al. 1993a\cite{Sa93a}; Bird 1994\cite{Bi94};
Escalera et al. 1994\cite{Es94}; Serna \& Gerbal 1996\cite{SG96};
Girardi et al. 1997\cite{Gi97}; Gurzadyan \& Mazure
1998\cite{GM98}). Consensus on the results, however, has been
frequently hindered by differences on the definition of substructure
adopted, on the methodology applied, on the scale used to examine the
spatial distribution of the galaxies, and even on the levels of
significance chosen to discriminate between real structure and
statistical fluctuations.

The debate on the existence of substructure in clusters has been also
fueled by the lack of adequate cluster samples to look at the
problem. Optical datasets (we will not discuss here X-ray data) which
combine both positional and velocity information are essential to
determine unambiguously cluster membership and, hence, to eliminate
projection uncertainties on the evaluation of subclustering. On the
other hand, meaningful estimates of the amount of substructure within
rich clusters of galaxies require large catalogs of these systems, free
from sampling biases and representative of the total
population. Fortunately, a great deal of progress is now being made in
this direction thanks to the rapid development of multi-object
spectroscopy, which has made possible the emergence of extensive
redshift surveys of galaxies in clusters (e.g. Dressler \& Shectman
1988b\cite{DS88b}; Teague et al. 1990\cite{Te90}; Zabludoff et
al. 1990\cite{Za90}; Beers et al. 1991\cite{Be91}; Malumuth et
al. 1992\cite{Ma92}; Yee et al. 1996\cite{Ye96}).

The recently compiled ESO Nearby Abell Cluster Survey (ENACS) catalog
(Katgert et al. 1996\cite{Ka96}, 1998\cite{Ka98}) is the result of the
last and, by far, most extensive multi-object spectroscopic survey of
nearby rich clusters of galaxies. The survey was specifically designed
to provide good kinematical data for the construction, in combination
with literature data, of a large statistically complete volume-limited
sample of rich ACO (Abell, Corwin, \& Olowin 1989)\cite{ACO89} clusters
in a region of the sky around the South Galactic Pole (Mazure et
al. 1996)\cite{Ma96}. The catalog contains positions, isophotal (red)
R-magnitudes within the 25 mag arcsec$^{-2}$ isophote, and redshifts of
more than 5600 galaxies in the directions of 107 southern ACO clusters
with richness $R_{\mathrm{ACO}}\ge 1$ and mean redshifts $z\la
0.1$. More importantly, numerous ENACS systems offer the possibility of
extracting extended magnitude-limited galaxy samples with a good level
of completeness, which is essential for many aspects of the study of
the properties of rich clusters, in particular, for detecting
substructure.

In this paper, we investigate substructure in a large subset of the
ENACS cluster catalog formed by 67 well-sampled systems.  Previous
studies of subclustering in cluster samples of comparable size have
relied on matching separate datasets and thus could not attain a high
degree of homogeneity. We apply to our clusters a variety of well-known
and complementary statistical tests for substructure, which analyze
information from the projected positions of the galaxies and/or their
radial velocities. Our aim is to evaluate the fractions of clumpy ENACS
systems implied by the different techniques and to compare them with
results from former studies relying on the same substructure
diagnostics. We begin by discussing in Sect.~\ref{data} the selection
of our cluster sample. Subclustering is investigated in Sect.~\ref{kin}
by means of three powerful classical tests which examine the velocity
dimension of the cluster data. The moment-based coefficients of
skewness and kurtosis are used to detect deviations from Gaussianity in
the velocity distributions, which are often correlated with the
presence of substructure in galaxy clusters. We also apply the 3D
diagnostic for substructure defined in DS88, known as the $\Delta$
test, to search for localized spatial-velocity correlations. These
statistics are complemented in Sect.~\ref{a2p} by the two-point
correlation formalism (Salvador-Sol\'e et al. 1993b; hereafter
Sa93)\cite{Sa93b}, which is used to look for signs of small-scale
subclustering in the two dimensional galaxy
distributions. Section~\ref{summ} contains a summary and discussion of
our results.

\section{The cluster sample}\label{data}

A total of 220 compact redshift systems with at least 4 member galaxies
and redshifts up to $z\la 0.1$ have been identified in the ENACS
catalog by Katgert et al. (1996; see their Table~6)\cite{Ka96}. These
systems were defined by grouping all the galaxies separated by a gap of
less than 1000\,\kms\, from its nearest neighbor in velocity space
along the directions of the clusters targeted in the course of the
project. Membership for the systems with at least 50 galaxies in the
original compilation suffered further refinement through the removal of
interlopers (i.e. galaxies that are unlikely system members but that
were not excluded by the 1000 \kms\, fixed-gap criterion) by means of
an iterative procedure that relies on an estimate of the mass profile
of the system (see Mazure et al. 1996\cite{Ma96} for details).

The completeness (number of redshifts obtained vs number of galaxies
observed) of the ENACS data varies from one sample to another and as a
function of apparent magnitude. Katgert et al. (1998)\cite{Ka98} show
that the completeness of the entire catalog reaches a maximum of about
80\% at intermediate magnitudes and stays approximately constant up to
$R_{25}=17$. Most of the ENACS clusters have indeed its maximum
completeness (which oscillates between 60\% and 90\%) at about this
limit (Katgert et al. 1996)\cite{Ka96}. At the bright end, the
completeness decreases slightly due to the low central surface
brightness of some of the brightest galaxies with sizes larger than the
diameter of the Optopus fibers, while for $R_{25}\ga 17$ it falls
abruptly due to the smaller S/N-ratio of the spectra of the fainter
galaxies. According to these results, and in order to deal with galaxy
samples with the maximum level of completeness, we have removed from
the ENACS systems all the galaxies with an $R_{25}$ magnitude larger
than 17. Furthermore, to obtain minimally robust results we have
excluded from the present analysis those systems with less than 20
galaxies left after the trimming in apparent brightness. These
restrictions lead to a final cluster dataset of 67 compact redshift
systems with a good level of completeness in magnitude and
containing a minimum of 20 member galaxies each.

All but one (Abell 3559) of the 29 clusters for which several Optopus
plates were taken (within each plate spectroscopy was attempted only
for the 50 brightest galaxies) are included in our cluster
sample. These ``multiple-plate'' clusters identify the richest and more
compact in redshift space systems surveyed. One of these, the
``double'' cluster Abell 548, has been separated into its SW and NE
components (see e.g. Davis et al. 1995\cite{Da95}), hereafter referred
to as A0548W and A0548E, respectively. Our database also includes 3
large secondary systems seen in projection in the fields of two of the
29 multiple-plate clusters: the systems in the foreground and in the
background of Abell 151, designated here as A0151F and A0151B,
respectively, and the background galaxy concentration seen in the field
of Abell 2819, designated here as A2819B. The remaining 35 systems are
``single-plate'' clusters for which a unique Optopus field was defined
(they all have, then, $N\le 50$). These systems are identified in
tables and figures by an asterisk.

Detailed information about each one of the systems selected, 
including robust estimates of their main physical properties, can be
found along the series of ENACS papers, especially in the articles
cited in this section.

\section{Substructure diagnostics relying on velocity data}\label{kin}
\subsection{Description of the tests}
To detect deviations from Gaussianity in the cluster's velocity
distributions, we use the classical coefficients of skewness and
kurtosis, which have been shown to offer greater sensitivity than other
techniques based on the order statistics or the gaps of the datasets
(see e.g. Bird \& Beers 1993)\cite{BB93}. The coefficient of skewness,
which is the third moment about the mean, measures the asymmetry of the
distribution. It is computed as
\begin{equation}
S={1\over{\sigma}^3}\left[{1\over{N}}\sum_{i=1}^N(v_i-\vmean)^3\right]\;,
\end{equation}  
with $\vmean$ and $\sigma$ the mean velocity and standard deviation
determined from the observed line-of-sight velocities $v_i$ of the $N$
cluster members. A positive (negative) value of $S$ implies that the
distribution is skewed toward values greater (less) than the mean.

The kurtosis is the fourth moment about the mean and measures the
relative population of the tails of the distribution compared to its
central region. Since the kurtosis of a normal distribution is expected
to be equal to 3, the kurtosis coefficient is usually defined to be
neutrally elongated for a Gaussian, in the form
\begin{equation}
K={1\over{\sigma}^4}\left[{1\over{N}}\sum_{i=1}^N(v_i-\vmean)^4\right]-3\;.
\end{equation}
Positive values of $K$ indicate distributions peakier than 
Gaussian and/or with heavier tails, while negative values reflect boxy
distributions that are flatter than Gaussian and/or with lighter tails.
The significance of the empirical values of the above two coefficients is
simply given by the probability that they are obtained by chance in a
normal distribution.

\renewcommand{\baselinestretch}{1} 

\begin{table}
\caption[]{Results of the kinematical substructure tests}
\tableheadtwo
 A0013* &   37 & 0.386 & 0.024 & 0.035\\
 A0087* &   22 & 0.450 & 0.263 & 0.106\\
 A0118* &   28 & 0.332 & 0.197 & 0.201\\
 A0119 &    87 & 0.245 & 0.601 & 0.681\\
 A0151F &   23 &0.245 & 0.270 & 0.848\\
 A0151 &    42 &0.280 & 0.114 & 0.669\\
 A0151B &   21 &0.230 & 0.385 & 0.008\\
 A0168 &    74 &0.178 & 0.024 & 0.287\\
 A0229* &   23 &0.252 & 0.167 & 0.018\\
 A0295* &   26 &0.202 & 0.566 & 0.544\\
 A0367* &   23 &0.414 & 0.048 & 0.653\\
 A0514 &    63 &0.132 & 0.175 & 0.214\\
 A0548W &  109 &0.128 & 0.272 & $<$0.001\\
 A0548E &  100 &0.171 & 0.051 & $<$0.001\\
 A0754* &   39 &0.099 & 0.309 & 0.351\\
 A0957* &   34 &0.495 & 0.291 & 0.034\\
 A0978 &    57 &0.006 & 0.006 & 0.004\\
 A1069* &   35 &0.014 & 0.676 & 0.208\\
 A1809* &   30 &0.107 & 0.296 & 0.563\\
 A2040* &   37 &0.248 & 0.362 & 0.107\\
 A2048* &   23 &0.229 & 0.311 & 0.969\\
 A2052* &   35 &0.009 & 0.028 & 0.542\\
 A2401* &   23 &0.315 & 0.647 & 0.001\\
 A2569* &   30 &0.216 & 0.416 & 0.021\\
 A2717 &    28 &0.294 & 0.173 & 0.373\\
 A2734 &    45 &0.283 & 0.415 & 0.140\\
 A2755* &   22 &0.196 & 0.264 & 0.011\\
 A2799* &   36 &0.162 & 0.160 & 0.356\\
 A2800* &   32 &0.416 & 0.068 & 0.297\\
 A2819 &    40 &0.047 & 0.088 & 0.631\\
 A2819B &   36 &0.012 & 0.013 & 0.322\\
 A2854* &   22 &0.061 & 0.357 & 0.644\\
 A2911* &   22 &0.261 & 0.089 & 0.055\\
 A3093* &   20 &0.460 & 0.385 & 0.479\\
 A3094 &    46 &0.329 & 0.043 & 0.004\\
 A3111* &   35 &0.057 & 0.351 & 0.072\\
 A3112 &    67 &0.282 & 0.243 & 0.280\\
 A3122 &    62 &0.391 & 0.441 & 0.039\\
 A3128 &   152 &0.224 & 0.108 & $<$0.001\\
 A3151* &   29 &0.072 & 0.577 & 0.074\\
 A3158 &    95 &0.468 & 0.136 & 0.393\\
 A3194* &   32 &0.378 & 0.009 & 0.010\\
 A3202* &   27 &0.254 & 0.108 & 0.052\\
 A3223 &    64 &0.000 & 0.004 & 0.162\\
 A3341 &    48 &0.404 & 0.007 & 0.910\\
 A3354 &    48 &0.169 & 0.424 & $<$0.001\\
 A3365* &   28 &0.221 & 0.002 & 0.005\\
 A3528* &   28 &0.192 & 0.039 & 0.277\\
 A3558 &    40 &0.329 & 0.146 & 0.186\\
 A3562 &    52 &0.025 & 0.253 & 0.003\\
 A3651 &    78 &0.446 & 0.254 & 0.026\\
 A3667 &   102 &0.249 & 0.581 & 0.199\\
 A3691* &   31 &0.116 & 0.221 & 0.203\\
 A3695 &    67 &0.220 & 0.408 & $<$0.001\\
 A3705* &   22 &0.299 & 0.175 & 0.044\\
\noalign{\smallskip}
\hline
\end{tabular}
\end{flushleft}
\label{Table1}
\end{table}

\setcounter{table}{0}
\begin{table}
\caption[]{(continued)}
\tableheadtwo
 A3733* &  41 & 0.140 & 0.558 & 0.409\\
 A3744 &   59 & 0.022 & 0.166 & 0.153\\
 A3764* &  33 & 0.037 & 0.043 & 0.759\\
 A3806 &   97 & 0.020 & 0.222 & 0.058\\
 A3809 &   80 & 0.109 & 0.204 & 0.274\\
 A3822 &   68 & 0.117 & 0.268 & 0.038\\
 A3825 &   45 & 0.194 & 0.424 & 0.160\\
 A3864* &  32 & 0.328 & 0.576 & 0.935\\
 A3879 &   33 & 0.099 & 0.003 & 0.452\\
 A3921* &  32 & 0.221 & 0.022 & 0.767\\
 A4008* &  24 & 0.220 & 0.268 & 0.407\\
 A4010* &  27 & 0.259 & 0.515 & 0.930\\
\noalign{\smallskip}
\hline
\end{tabular}
\end{flushleft}
\label{Table1cont}
\end{table}

\renewcommand{\baselinestretch}{2} 

Together with the above normality tests, we apply also the $\Delta$
test of DS88, which is a simple and powerful 3D substructure diagnostic
designed to look for local correlations between galaxy positions and
velocity that differ significantly from the overall distribution within
the cluster. It is based on the comparison of a local estimate of the
velocity mean $\vmean_\loc$ and dispersion $\sigma_\loc$ for each
galaxy with measured radial velocity, with the values of these same
kinematical parameters for the entire sample. The presence of
substructure is quantified by means of a sole statistic defined from 
the sum of the local kinematic deviations $\di$ over the $N$ cluster
members, in the form (Bird 1994)\cite{Bi94}
\begin{eqnarray} \label{delta}  
\Delta & = & \sum_{i=1}^N\di\nonumber\\
& = & \sum_{i=1}^N\left[{{\mathrm{nint}}(\sqrt N)+1\over{\sigma^2}}\left((\vmean_{\loc,i}-\vmean)^2+(\sigma_{\loc,i}-\sigma)^2\right)\right]^{1\over{2}},
\end{eqnarray}
with ${\mathrm{nint}}(x)$ standing for the integer nearest to $x$.
To avoid the formulation of any hypothesis on the form of the velocity
distribution of the parent population, the $\Delta$ statistic is
calibrated by means of Monte-Carlo simulations (we perform 1000 per
cluster) that randomly shuffle the velocities of the cluster members
while keeping their observed positions fixed. In this way any existing
local correlation between velocities and positions is destroyed. The
probability of the null hypothesis that there are no such correlations
is given in terms of the fraction of simulated clusters for which their
cumulative deviation is smaller than the observed value.

\subsection{Results}\label{kin_res}

Table~1 summarizes, for each one of the 67 magnitude-limited galaxy
samples defined in Sect.~\ref{data}, the number of galaxies $N$ meeting
the selection criteria and the probabilities that the empirical values
of the three statistics described above could have arisen by chance
(the smaller the quoted value the more significant is the departure
from the null hypothesis). At the 5\% significance level (in this
section all results will be referred to this level of significance)
about 30\% (20 of 67) of the systems exhibit a non-Gaussian velocity
distribution according to at least one of the two normality tests. This
is a little small fraction if compared with the results of previous
studies by West \& Bothun (1990)\cite{WB90}, Bird \& Beers
(1993)\cite{BB93}, and Bird (1994)\cite{Bi94}, in which $\sim40\%-50\%$
of the clusters investigated had radial velocity distributions with
non-normal values of the skewness and/or kurtosis. The discrepancies,
however, are not statistically significant and point to possible biases
towards the inclusion of clumpy systems in former cluster datasets
(see, for instance, the selection criteria applied by Dressler \&
Shectman 1988b\cite{DS88b}).  The normality tests do not detect either
significant differences between the single- and multiple-plate subsets,
which indicate frequencies of rejection of the Gaussian hypothesis,
26\% (9/35) and 34\% (11/32) respectively, fully compatible within the
statistical uncertainties.

On the other hand, 31\% (21/67) of our clusters are found to
show substructure according to the $\Delta$ test. In a recent
investigation of the kinematics and spatial distribution of the
Emission-Line Galaxies (ELG) in clusters, Biviano et
al. (1997)\cite{Bi97} have applied this same test to the 25 ENACS
systems with $N\ge 50$ that contain at least one ELG, finding evidence
for substructure in $\sim40$\% of the cases. As was to be expected,
this value is in excellent agreement with the 38\% (12/32) of the
multiple-plate systems which demonstrate substructure in our
dataset. Previous analysis of subclustering also relying on the
$\Delta$ statistic by Escalera et al. (1994)\cite{Es94} and Bird
(1994)\cite{Bi94} claim similar percentages of clumpy systems, 38\%
(6/16) and 44\% (11/25) respectively, while the fraction quoted in the
original work by DS88 is somewhat higher, 53\% (8/15), but still
compatible with the other results within the statistical
uncertainties. We emphasize, however, that none of the preceding works
payed attention to the completeness in magnitude of the galaxy samples
under scrutiny. As in the case of the Gaussianity tests, we do not find
significant differences between the fractions of substructured
single-plate systems (9/35) and multiple-plate ones (12/32) indicated
by the $\Delta$ statistic.

\renewcommand{\baselinestretch}{1} 

\begin{table*}
\caption[]{Characteristics of the 67 clusters}
\tableheadone
 A0013* & 00 & 11 & 03$\fs$0 & $-$19 & 47 & 40 & 0.17 & $-$0.26 & 0.42 &  20\\
 A0087* & 00 & 40 & 15$\fs$8 & $-$10 & 05 & 01 & 0.80 &  0.57 & 0.54 &  20\\
 A0118* & 00 & 52 & 45$\fs$6 & $-$26 & 38 & 27 & 0.80 &  0.83 & 1.17 &  28\\
 A0119 &  00 & 53 & 39$\fs$6 & $-$01 & 30 & 39 & 0.84 &  1.39 & 0.92 &  81\\
 A0151F & 01 & 06 & 24$\fs$3 & $-$16 & 14 & 50 & 0.58 & $-$0.94 & 0.19 &  8\\
 A0151 &  01 & 06 & 44$\fs$4 & $-$15 & 45 & 25 & 0.32 & $-$0.48 & 0.68 &  33\\
 A0151B & 01 & 06 & 12$\fs$3 & $-$15 & 51 & 07 & 0.50 &  0.69 & 1.25 &  17\\
 A0168 &  01 & 12 & 31$\fs$0 &  00 & 02 & 26 & 0.49 & $-$1.07 & 0.76 &  62\\
 A0229* & 01 & 37 & 03$\fs$5 & $-$03 & 55 & 14 & 0.39 & $-$0.33 & 0.51 &  11\\
 A0295* & 01 & 59 & 34$\fs$5 & $-$01 & 19 & 20 & 0.80 &  0.32 & 0.36 &  21\\
 A0367* & 02 & 34 & 23$\fs$6 & $-$19 & 33 & 59 & 0.64 & $-$0.55 & 0.89 &  19\\
 A0514 &  04 & 45 & 56$\fs$8 & $-$20 & 35 & 49 & 0.58 & $-$0.37 & 0.97 &  56\\
 A0548W & 05 & 42 & 53$\fs$0 & $-$25 & 55 & 35 & 0.65 &  0.74 & 0.90 &  82\\
 A0548E & 05 & 46 & 00$\fs$9 & $-$25 & 32 & 32 & 0.96 &  0.49 & 1.12 &  98\\
 A0754* & 09 & 06 & 19$\fs$7 & $-$09 & 25 & 47 & 0.49 & $-$0.23 & 0.31 &  21\\
 A0957* & 10 & 11 & 09$\fs$5 &  00 & 38 & 47 & 0.22 &  0.07 & 0.23 &  23\\
 A0978 &  10 & 18 & 00$\fs$5 & $-$06 & 21 & 40 & 0.27 & $-$1.40 & 0.78 &  48\\
 A1069* & 10 & 37 & 13$\fs$9 & $-$08 & 21 & 54 & 0.40 & $-$1.43 & 0.46 &  25\\
 A1809* & 13 & 50 & 31$\fs$4 &  05 & 23 & 06 & 0.53 &  1.01 & 0.72 &  28\\
 A2040* & 15 & 10 & 24$\fs$2 &  07 & 36 & 58 & 0.88 & $-$0.13 & 0.44 &  31\\
 A2048* & 15 & 12 & 44$\fs$0 &  04 & 33 & 15 & 0.60 & $-$1.51 & 0.75 &  20\\
 A2052* & 15 & 14 & 23$\fs$7 &  07 & 15 & 25 & 0.73 &  1.40 & 0.33 &  28\\
 A2401* & 21 & 55 & 50$\fs$0 & $-$20 & 17 & 58 & 0.91 &  0.40 & 0.41 &  19\\
 A2569* & 23 & 15 & 09$\fs$9 & $-$13 & 06 & 12 & 0.95 & $-$1.12 & 0.66 &  22\\
 A2717 &  00 & 00 & 05$\fs$3 & $-$36 & 08 & 00 & 0.46 & $-$0.12 & 0.67 &  20\\
 A2734 &  00 & 08 & 46$\fs$1 & $-$29 & 06 & 36 & 0.52 &  0.03 & 0.90 &  37\\
 A2755* & 00 & 15 & 08$\fs$7 & $-$35 & 25 & 48 & 0.60 & $-$0.14 & 0.78 &  15\\
 A2799* & 00 & 34 & 55$\fs$9 & $-$39 & 27 & 15 & 0.48 &  0.57 & 0.67 &  32\\
 A2800* & 00 & 35 & 35$\fs$6 & $-$25 & 24 & 26 & 0.51 &  1.12 & 0.44 &  18\\
 A2819 &  00 & 43 & 40$\fs$5 & $-$63 & 49 & 35 & 0.31 &  0.26 & 1.30 &  33\\
 A2819B & 00 & 43 & 38$\fs$5 & $-$63 & 50 & 59 & 0.54 & $-$0.15 & 2.05 &  34\\
 A2854* & 00 & 58 & 23$\fs$1 & $-$50 & 46 & 39 & 0.40 & $-$1.05 & 0.43 &  14\\
 A2911* & 01 & 23 & 57$\fs$4 & $-$38 & 11 & 37 & 0.65 & $-$0.50 & 0.33 &  10\\
 A3093* & 03 & 09 & 18$\fs$1 & $-$47 & 35 & 55 & 0.19 & $-$0.23 & 0.55 &  12\\
 A3094 &  03 & 09 & 51$\fs$8 & $-$27 & 09 & 29 & 0.38 & $-$0.48 & 0.99 &  39\\
 A3111* & 03 & 15 & 49$\fs$6 & $-$45 & 52 & 04 & 0.69 & $-$0.84 & 0.73 &  26\\
 A3112 &  03 & 16 & 18$\fs$6 & $-$44 & 27 & 07 & 0.88 &  0.40 & 1.60 &  65\\
 A3122 &  03 & 20 & 35$\fs$8 & $-$41 & 30 & 45 & 0.65 &  0.14 & 0.86 &  39\\
 A3128 &  03 & 28 & 52$\fs$7 & $-$52 & 48 & 57 & 0.61 &  0.45 & 2.19 & 152\\
 A3151* & 03 & 38 & 30$\fs$7 & $-$28 & 51 & 41 & 0.28 &  0.11 & 0.31 &  22\\
 A3158 &  03 & 41 & 13$\fs$6 & $-$53 & 47 & 53 & 0.73 &  0.21 & 1.34 &  77\\
 A3194* & 03 & 57 & 04$\fs$4 & $-$30 & 19 & 01 & 0.23 &  1.12 & 0.39 &  14\\
 A3202* & 03 & 59 & 30$\fs$0 & $-$53 & 48 & 56 & 0.38 & $-$0.01 & 0.56 &  19\\
 A3223 &  04 & 06 & 16$\fs$7 & $-$31 & 02 & 09 & 0.57 &  1.25 & 1.02 &  53\\
 A3341 &  05 & 23 & 44$\fs$2 & $-$31 & 34 & 58 & 0.70 &  1.30 & 0.70 &  45\\
 A3354 &  05 & 32 & 45$\fs$0 & $-$28 & 36 & 08 & 0.43 & $-$1.23 & 1.09 &  43\\
 A3365* & 05 & 46 & 07$\fs$9 & $-$21 & 55 & 57 & 0.34 &  0.53 & 0.59 &  22\\
 A3528* & 12 & 51 & 44$\fs$1 & $-$28 & 45 & 03 & 0.84 &  0.91 & 0.33 &  15\\
 A3558 &  13 & 25 & 49$\fs$7 & $-$31 & 13 & 26 & 0.62 & $-$0.89 & 0.85 &  34\\
 A3562 &  13 & 28 & 28$\fs$0 & $-$31 & 26 & 09 & 0.55 & $-$0.69 & 0.57 &  23\\
 A3651 &  19 & 48 & 15$\fs$1 & $-$55 & 12 & 31 & 0.40 & $-$0.10 & 1.12 &  43\\
 A3667 &  20 & 07 & 53$\fs$8 & $-$56 & 56 & 59 & 0.34 & $-$0.61 & 1.50 &  87\\
 A3691* & 20 & 31 & 02$\fs$5 & $-$38 & 12 & 57 & 0.55 & $-$1.36 & 0.64 &  21\\
\noalign{\smallskip}
\hline
\end{tabular}
\end{flushleft}
\label{Table2}
\end{table*}

\setcounter{table}{1}
\begin{table*}
\caption[]{(continued)}
\tableheadone
 A3695 &  20 & 31 & 39$\fs$8 & $-$36 & 00 & 12 & 0.47 & $-$1.19 & 1.34 &  47\\
 A3705* & 20 & 38 & 40$\fs$0 & $-$35 & 23 & 55 & 0.23 &  0.36 & 0.45 &  13\\
 A3733* & 20 & 58 & 52$\fs$3 & $-$28 & 18 & 41 & 0.29 & $-$1.50 & 0.28 &  19\\
 A3744 &  21 & 04 & 22$\fs$6 & $-$25 & 41 & 35 & 0.83 &  0.34 & 0.62 &  46\\
 A3764* & 21 & 22 & 58$\fs$4 & $-$35 & 01 & 09 & 0.09 & $-$1.15 & 0.24 &  11\\
 A3806 &  21 & 41 & 38$\fs$2 & $-$57 & 24 & 42 & 0.31 & $-$0.26 & 1.85 &  79\\
 A3809 &  21 & 44 & 02$\fs$8 & $-$44 & 10 & 58 & 0.40 & $-$0.12 & 0.97 &  59\\
 A3822 &  21 & 50 & 22$\fs$7 & $-$58 & 03 & 14 & 0.40 & $-$0.13 & 2.05 &  61\\
 A3825 &  21 & 54 & 45$\fs$8 & $-$60 & 37 & 07 & 0.54 & $-$0.32 & 1.26 &  34\\
 A3864* & 22 & 16 & 50$\fs$4 & $-$52 & 45 & 28 & 0.34 & $-$0.53 & 1.13 &  26\\
 A3879 &  22 & 24 & 00$\fs$0 & $-$69 & 14 & 03 & 0.55 & $-$0.11 & 0.93 &  24\\
 A3921* & 22 & 46 & 30$\fs$6 & $-$64 & 40 & 33 & 0.39 &  0.00 & 1.43 &  27\\
 A4008* & 23 & 27 & 36$\fs$9 & $-$39 & 34 & 04 & 0.76 &  1.20 & 0.52 &  18\\
 A4010* & 23 & 28 & 49$\fs$6 & $-$36 & 47 & 31 & 0.42 &  0.36 & 0.72 &  22\\
\noalign{\smallskip}
\hline
\end{tabular}
\end{flushleft}
\label{Table2cont}
\end{table*}

\renewcommand{\baselinestretch}{1}

\section{The average two-point correlation function}\label{a2p}
\subsection{Definition and practical implementation}
The average two-point correlation function $\bar \xi$ (see Sa93 for
details) was introduced for the statistical characterization of
subclustering in inhomogeneous systems with isotropy around one single
point. Given a circularly symmetric galaxy cluster, this statistic can
be calculated exactly as the usual two-point correlation function in
the homogeneous and isotropic case through the expression
\begin{equation}\label{a2pcf}  
\bar \xi(s)={(\rho\ast\rho)(s)-(n\ast n)(s)\over{(n\ast n)(s)}}-{1\over{N}}\;,
\end{equation}
with $\rho$ some continuous function approximating the observed number
density distribution of galaxies, and $n$ the mean radial number
density profile estimated from the azimuthal average of $\rho$. Notice
that, contrarily to $\rho$, $n$ is insensitive to the existence of
correlation in galaxy positions. The additive constant $1/N$ in
Eq.~(\ref{a2pcf}) corrects for the negative bias caused by the fact
that each cluster galaxy chosen at random has {\em{only}} $N-1$
neighbors, one less than the number expected for a fully random
process.

\begin{figure*}
  \resizebox{\hsize}{!}{\includegraphics{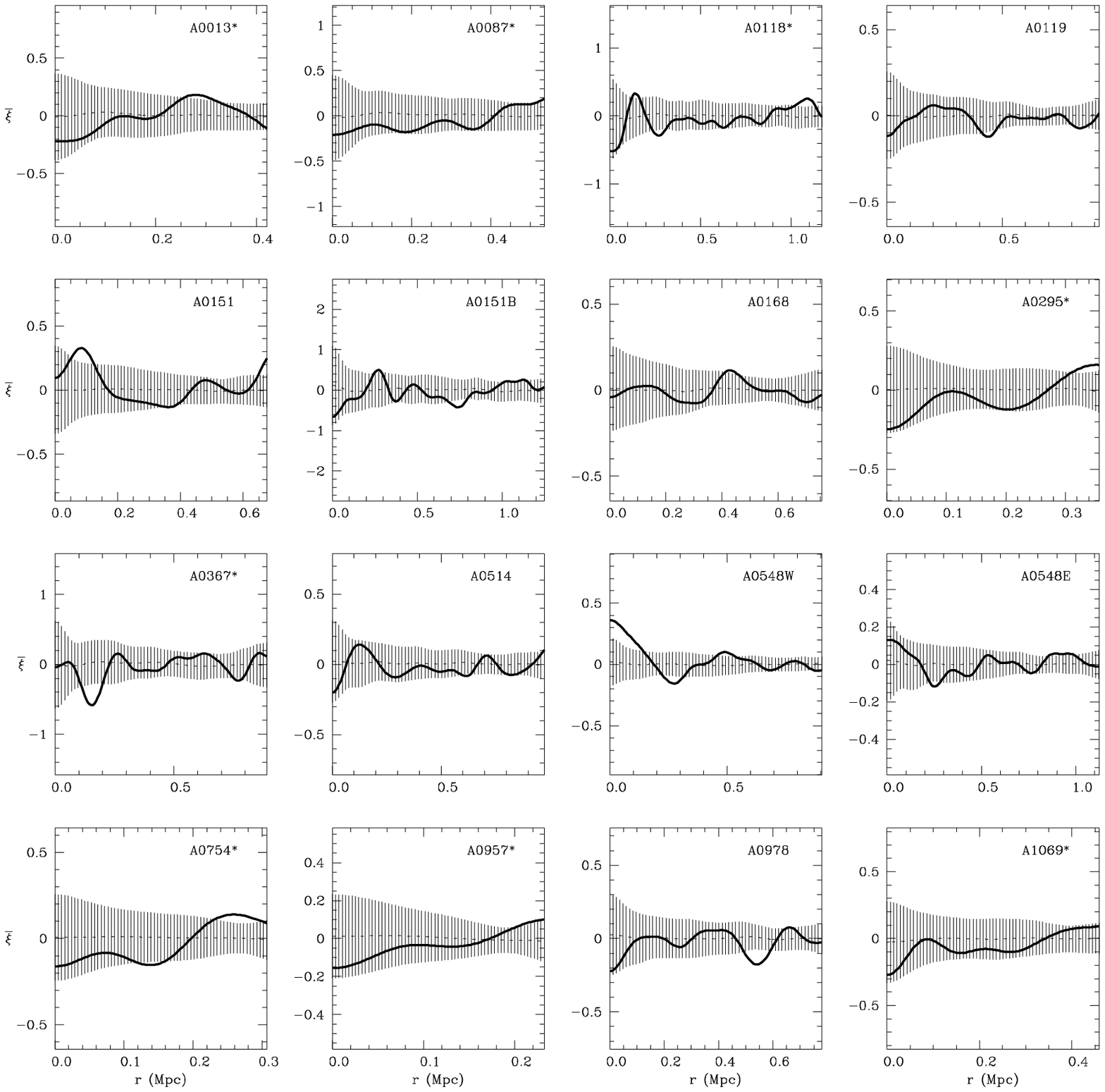}} \vspace{-1.5cm}
  \caption{Average two-point correlation function for the 59 circularly
  symmetric galaxy samples with $N_{\mathrm c}\ge 15$ compared with the
  mean solution (dashed line) and $1\sigma$-error (vertical solid
  lines) obtained from 200 Poissonian simulations of each cluster. The
  spatial resolution is 0.05\,\mpc.}  \label{Fig. 1_1}
\end{figure*}

\setcounter{figure}{0}
\begin{figure*}
  \resizebox{\hsize}{!}{\includegraphics{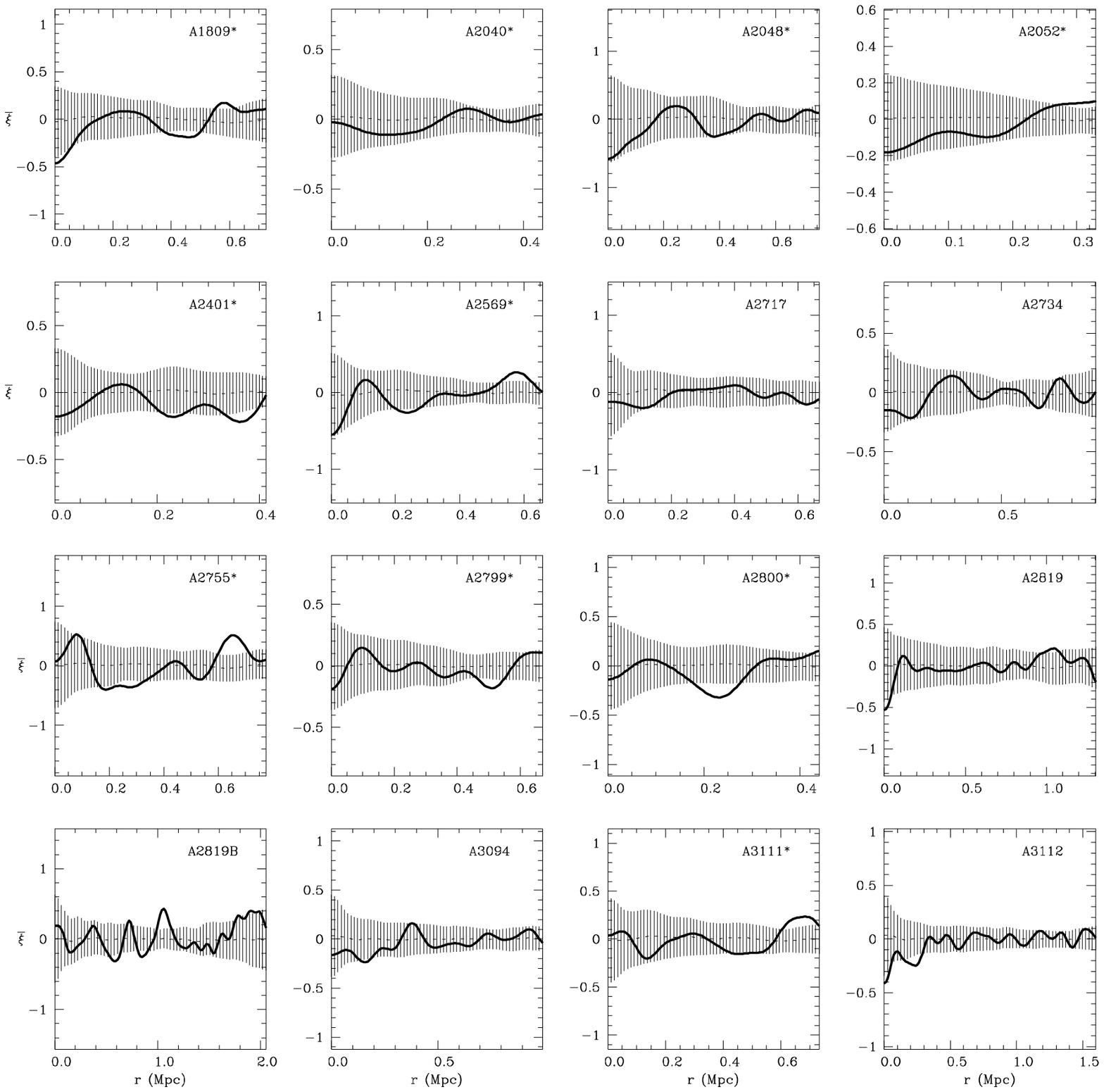}}
  \vspace{-1.5cm}
  \caption{(continued)}
  \label{Fig. 1_2}
\end{figure*}

\setcounter{figure}{0}
\begin{figure*}
  \resizebox{\hsize}{!}{\includegraphics{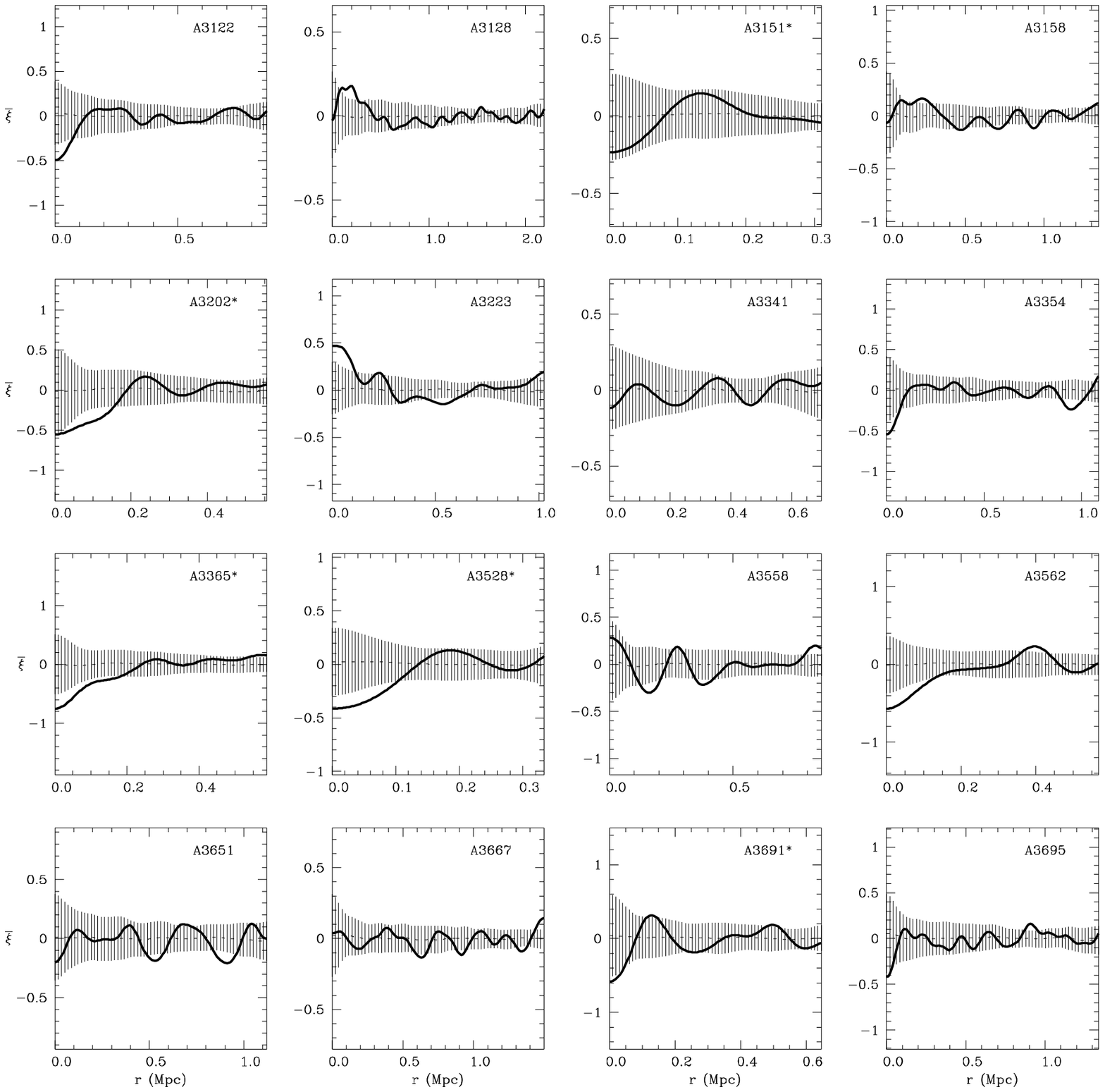}}
  \vspace{-1.5cm}
  \caption{(continued)}
  \label{Fig. 1_3}
\end{figure*}

\setcounter{figure}{0}
\begin{figure*}
  \resizebox{\hsize}{!}{\includegraphics{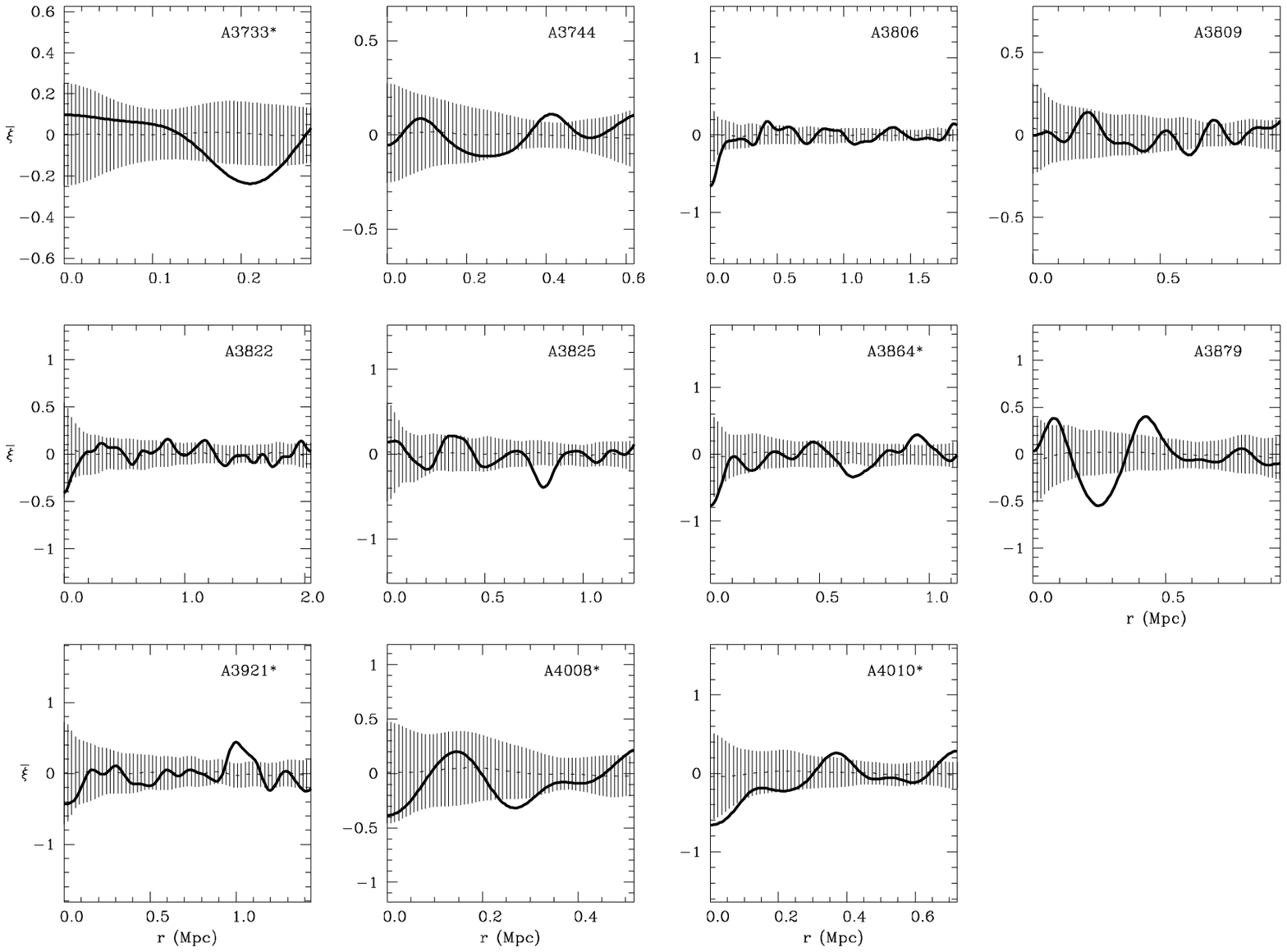}}
  \vspace{-5.4cm}
  \caption{(continued)}
  \label{Fig. 1_4}
\end{figure*}

The autocorrelation products $\rho\ast\rho$ and $n\ast n$ are computed
via the sequence of transformations (see also Sal\-va\-dor-So\-l\'e et
al. 1993a)\cite{Sa93a}
\begin{equation}\label{ndec2}  
(\rho\ast\rho)(s)={{\cal F}_1\circ {\cal A}}\biggl[{{\cal A}\circ 
{\cal F}_{1}^{-1}}\biggl( 2\int_{s}^{\infty}P(x)\,\der x\biggr)\biggr]\;,
\end{equation}
and
\begin{equation}\label{ndir2} 
(n\ast n)(s)={{\cal F}_1\circ {\cal A}}\biggl[{{\cal A}\circ 
{\cal F}_{1}^{-1}}\biggl(\int_{s}^{\infty}\Pi(x)\,\der x\biggr)\biggr]^{2}\;,
\end{equation} 
which rely, respectively, on the calculation of the cumulative forms of
$P(s)\,\der s$, the number of pairs of galaxies with observed
separation between $s$ and $s + \der s$ among the $N(N-1)/2$ galaxy
pairs obtained from the cluster sample, and of $\Pi(s)\,\der s$, the
number of galaxies at projected distances between $s$ and $s + \der s$
from the center of symmetry of the galaxy distribution. In
Eqs.~(\ref{ndec2}) and (\ref{ndir2}) ${\cal F}_1$ and $\cal A$ stand,
respectively, for the one-dimensional Fourier and Abel transformations,
while the symbol ``$\circ$'' denotes the composition of functions. From
the latter two equations it is readily apparent that the numerical
estimate of $\bar \xi$ is independent of the bin size used for the
integrals $\int_{s}^{\infty} P(x)\,\der x$ and
$\int_{s}^{\infty}\Pi(x)\,\der x$, which merely determines the sampling
interval of the solution, so there are no lower limits on the size of
the subclumps that can be detected (nor a priori assumptions on their
possible number and shapes are required). Nonetheless, it is advisable
to attenuate the statistical noise of $\bar\xi(s)$ at galactic scales
(Sa93). Thus, we apply a low-passband hamming filter leading to a final
resolution length of 0.05\,\mpc. Notice also that the use of the
cumulative forms of the distributions $P(s)$ and $\Pi(s)$ makes this
statistic particularly well suited for galaxy samples containing a
small number of objects.

The statistical significance of substructure for each cluster is
obtained by checking the null hypothesis that the observed $\rho(s)$
arises from a Poissonian realization of an unknown theoretical density
profile, which is approximated by $n(s)$. In practice, this translates
to a comparison between the empirical function given by
Eq.~(\ref{a2pcf}) with the mean and one standard deviation of the same
function obtained from a large number of Poissonian cluster simulations
(i.e. both the radius and the azimuthal angle of each galaxy are
chosen at random) that reproduce the profile $n(s)$.

\subsection{Results}\label{a2p_res}
In order to apply this diagnostic, circularly symmetric galaxy
subsamples have been extracted from our dataset by means of a
three-step procedure. The first step consists in the determination of
the system barycenter through an iterative process that uses only those
galaxies located inside the maximum circle, around the centroid
obtained in the previous iteration, inscribed within the limits of the
surveyed field. This procedure mitigates any incompleteness in position
caused by the spatial filters used in the data acquisition and, when
several structures are present in the same region, tends to focus on
the main subsystem. A second iterative process calculates the system
ellipticity $e$, which is assumed to be homologous, and the orientation
$\theta$ of its major axis. Analogously to the barycenter
determination, galaxies located in incomplete (elliptical) spatial
bins around the barycenter are excluded from the calculations. Finally,
circular symmetry is ensured by contracting the galaxy coordinates
along the semimajor axis by $\sqrt{e}$ and expanding the semiminor-axis
coordinates by the inverse of this same factor. In this manner, we take
into account elongation effects that might artificially indicate
clumpiness, while any true signal of subclustering is preserved.

Table~2 lists, system by system, the barycenter coordinates, the values
of the parameters $e$ and $\theta$ (relative to the WE
direction)\footnote{Adami et al. (1998)\cite{Ad98} have also inferred
these parameters for a number of clusters in this list from
Maximum-Likelihood fits to the COSMOS data, obtaining compatible
results.}, the equivalent radius $r_{\mathrm eq}$ (i.e. the radius of a
circle with an area equal to the maximum elliptical isopleth contour)
in \mpc, and the number of galaxies $N_{\mathrm c}$ included in the
circularly symmetric subsamples. Physical units have been inferred from
the cosmological distances of the clusters, which are calculated by
correcting their mean heliocentric redshifts to the Cosmic Microwave
Background rest frame according to the dipole measured by Kogut et
al. (1993)\cite{Ko93}.

\begin{figure} 
  \resizebox{\hsize}{!}{\includegraphics{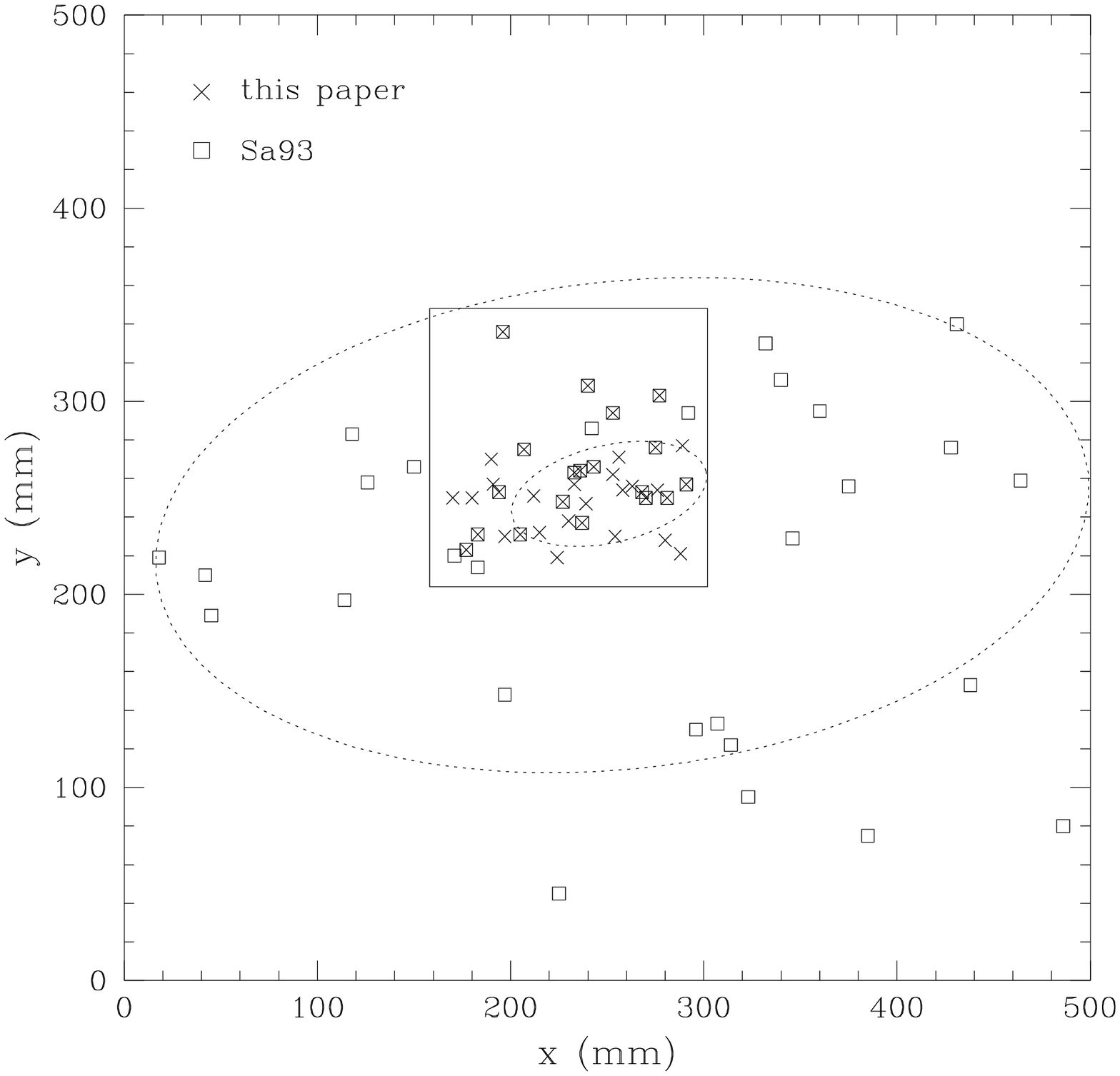}}
  \caption{Superposition of the galaxy plots corresponding to the
  magnitude-limited samples of the cluster Abell 754 defined in the
  present study and in Sa93. Likely members of the two datasets are
  identified by diagonally-crossed open squares. The inner square
  delimits the area used here for the ENACS data. The dotted ellipses
  inscribed within each field encompass the objects that participate in
  the calculation of $\bar\xi(s)$. Coordinates are in millimeters
  (scale $10\farcs9\,\mbox{mm}^{-1}$).}  \label{Fig. 2}
\end{figure}

To minimize small-number effects, the calculation of the average
two-point correlation function was restricted to the 59 circularly
symmetric galaxy subsamples with 15 or more objects. The results are
depicted in Fig.~1, together with the mean solutions and
$1\sigma$-errors resulting from 200 Poissonian realizations of each
cluster. These plots show that only 6 systems, A0151, A0548W, A2755*,
A3128, A3223, and A3879, have a strictly positive signal raising above
the noise at separations smaller than 0.2\,\mpc\ (as in Sa93, we
consider the presence of central maxima reaching at least the $1\sigma$
level as indicative of small-scale subclustering). Two other systems,
A0118* and A3691*, exhibit also a positive departure of more than
$1\sigma$ at these small scales, but have negative central values of
$\bar\xi$. Indeed, about three fourths (46 of 59) of the clusters in
our sample present negative central signals which, in 15 cases, even go
over $1\sigma$. 

These results are in notorious contradistinction with those inferred in
Sa93 from the analysis of 14 of the 15 Dressler \& Shectman's
(1988b)\cite{DS88b} clusters (Abell 548 was excluded). In this earlier
study {\em{all}} systems gave positive central values of $\bar\xi$ and
nine showed departures between 1 and $2\sigma$ at separations inferior
to 0.2\,\mpc. The only cluster in common between both investigations,
Abell 754, is found here to exhibit no evidence for substructure, yet
in Sa93 this cluster was seen to produce, with a similar number of
objects, a strong positive central signal. One plausible origin of that
conflict could be the very different areal coverage of the galaxy
samples used in the two studies for this particular cluster (see
Fig.~2; but notice that the orientations, ellipticities and barycenter
positions are, nevertheless, in very good agreement). This would be the
case if the positive detection in Sa93 was produced by small subgroups
located outside the cluster core. We also point out the suggestion made
in Sa93 that the asymmetry shown by the projected galaxy distribution
in the Dressler \& Shectman's field, not noticeable at short distances
from the cluster center, could have caused the observed signal.

It is interesting to note that similarly strong discrepancies can be
observed with respect to the results of the wavelet analysis of
substructure performed by Escalera et al. (1994)\cite{Es94}. These
authors found that only three systems among the 16 that they
investigated, most of them Dressler \& Shectman's clusters, did not
show significant small-scale subclustering. Further, there is a good
agreement between the results of this study and of Sa93 for the common
clusters.

Finally, it is also worth noting that, with the exception of the system
A0151, the remaining five clusters with evidence for small-scale
structure in the galaxy positional data show also signs of
substructure in velocity space (see Table~1).

\section{Summary and discussion}\label{summ}
We have evaluated here the frequency of subclustering in 67
well-sampled nearby rich galaxy clusters extracted from the list of 220
compact redshift systems identified in the homogeneous ENACS
catalog. Three classical diagnostics sensitive to correlations in
velocity space have registered amounts of substructure comparable with
those found in earlier studies which applied the same estimators to
datasets less representative of the nearby rich cluster population. The
average two-point correlation function statistic has allowed us to
investigate the clumpiness of the two dimensional galaxy distributions
at small intergalactic separations. In doing so we have found that only
about one of every 10 systems studied shows evidence for positive
correlation among the projected positions of its member galaxies at
scales inferior to $0.2$\,\mpc. This result contrasts markedly with the
very high fraction of Dressler \& Shectman's clusters which
demonstrated signs of small-scale substructure in the earlier analysis
by Sa93 (see also Escalera et al. 1994).\cite{Es94}

It is possible that already mentioned factors such as cluster selection
biases, likely affecting some of the existing catalogs, or the
restricted coverage of the galaxy distributions of part of the clusters
studied here may be partially responsible for this conflicting
result. Nevertheless, there are grounds for believing that it could be
caused too by an increase of the incompleteness of the ENACS galaxy
samples at small scales. A telling argument in support of this latter
viewpoint is that 25 of the 28 magnitude-limited single-plate systems
(and 21 of the 31 multiple-plate ones) for which the $\bar\xi$
statistic has been inferred exhibit negative central values of this
function.  Since in the absence of correlation among galaxy positions
positive and negative values of $\bar\xi(0)$ are equally probable (see
Sa93), we infer that the ENACS clusters do show suggestive evidence of
a systematic deficiency of galaxies at very short separations.

What then could have originated this effect? Let us remember that the
ENACS project was aimed to obtain extensive redshift data in the fields
of more than 100 rich galaxy clusters. To achieve this goal in a
reasonable amount of time the number of exposures for each targeted
cluster was minimized, making it difficult to compensate the
operational restrictions inherent to the fiber-optic system
(limited number of fibers available, minimum distance allowed for
the positioning of contiguous fibers, etc) with redundant exposures.
This might well have affected the reproduction of the finest details of
the cluster galaxy distributions, especially when the coverage was done
by means of a single plate. Notice, in this regard, that only one
single-plate cluster is among the 6 systems that show signs of
small-scale substructure in our dataset. (One may wonder if this latter
result could have been produced instead by the relatively small galaxy
populations of the single-plate clusters; this possibility is
challenged, however, by the fact that in Sa93 seven of the 9 systems
which gave a positive detection had less than 50 objects.) The reduced
success in the redshift measurement for the brightest galaxies (see
Sect.~\ref{data}, and Katgert et al. 1996)\cite{Ka96}, which are fair
tracers of substructures within clusters (Biviano et
al. 1996\cite{Bi96}; Gurzadyan \& Mazure 1998\cite{GM98}) is another
factor which may have also contributed to conceal the presence of small
subgroups.

To sum up, the amount of intermediate and large-scale subclustering
detected in the ENACS systems is in fair agreement with, and therefore
validates, the results of previous analysis of substructure in nearby
rich clusters based on less homogeneous datasets. The present
investigation, however, has revealed that the ENACS galaxy samples
could suffer from an increasing incompleteness towards small
intergalactic separations. In this regard, we caution that the ENACS
data by themselves may be insufficient in applications requiring a
detailed description of the small-scale substructuring properties of
clusters, such as those that investigate the formation of these systems
and its consequences on cosmological theories.

\begin{acknowledgements}
The authors would like to thank Peter Katgert for enlightening and
useful discussions. GGC acknowledges support by the Universitat
Polit\`ecnica de Catalunya, through research contract PR97--07. This
work has been supported by the Direcci\'on General de Investigaci\'on
Cient\'{\i}fica y T\'ecnica, under contract PB96--0173.
\end{acknowledgements}

\end{document}